\documentclass{PoS}

\title{Magnetic fields in nearby galaxies}

\ShortTitle{Magnetic fields in nearby galaxies}

\author{\speaker{George Heald}\\
        The Netherlands Foundation for Radio Astronomy (ASTRON)\\
        E-mail: \email{heald@astron.nl}}

\author{Robert Braun\\
        CSIRO-ATNF\\
        E-mail: \email{Robert.Braun@csiro.au}}

\abstract{We describe a recent full-polarization radio continuum survey, performed using the Westerbork Synthesis Radio Telescope (WSRT), of several nearby galaxies in the Spitzer Infrared Nearby Galaxies Survey (SINGS) sample. The WSRT-SINGS survey has been utilized to study the polarized emission and Faraday rotation measures (RMs) in the targets, and reveals an important new observational trend. The azimuthal distribution of polarized flux seems to be intimately related to the kinematic orientation of galaxies, such that in face-on galaxies the lowest level of polarized flux is detected along the kinematic major axis. In highly inclined galaxies, the polarized flux is minimized on both ends of the major axis, and peaks near the minor axis. Using models of various three-dimensional magnetic field geometries, and including the effects of turbulent depolarization in the midplane, we are able to reproduce the qualitative distribution of polarized flux in the target galaxies, its variation with inclination, and the distribution of RMs, thereby constraining the global magnetic field structure in galaxies. Future radio telescope facilities, now being planned and constructed, will have properties making them extremely well-suited to perform vastly larger surveys of this type, and are thereby poised to significantly increase our understanding of the global structure of galactic magnetic fields. We discuss progress that can be made using surveys which will be realized with these new facilities, focusing in particular on the Aperture Tile in Focus (APERTIF) and Australian Square Kilometre Array Pathfinder (ASKAP) telescopes, both based on Focal Plane Array (FPA) designs, which are expected to be particularly useful for wide-field polarization applications.}

\FullConference{Panoramic Radio Astronomy: Wide-field 1-2 GHz research on galaxy evolution - PRA2009\\
		 June 02 - 05 2009\\
		 Groningen, the Netherlands}

\begin{document}

\section{The WSRT-SINGS Survey}

Magnetic fields are a critical component of the interstellar medium (ISM) in galaxies, yet major questions still remain about their properties. One of the unknowns is the three-dimensional structure of galactic magnetic fields. The most effective way to measure magnetic fields in galaxies is through radio measurements of synchrotron emission (using the polarized emission to study the ordered fields perpendicular to the line of sight, and its associated Faraday rotation to trace fields parallel to the line of sight). Observations of face-on spiral galaxies have shown that the planar component of magnetic fields traces the spiral arms (e.g. \cite{beck_2009}), while observations of edge-on disks tend to show a characteristic {\sf X}-shaped morphology (e.g. \cite{krause_2008}). Galactic magnetic fields are expected to take either a dipolar or quadrupolar form (e.g. \cite{widrow_2002}), but previous observations of external galaxies have not been able to strongly constrain which type of field might be present.

The Westerbork Synthesis Radio Telescope (WSRT) was recently used to observe a sample of several Spitzer Infrared Nearby Galaxies Survey (SINGS; \cite{kennicutt_etal_2003}) galaxies, in full polarization, in order to provide supplementary data on the radio continuum emission. The WSRT-SINGS survey itself is described by \cite{braun_etal_2007}, and the polarization analysis by \cite{heald_etal_2009}. Briefly, each of the targets was observed for a total of 12 hours. Polarization data were obtained in two wide bands (effective bandwidth $\approx132\,\mathrm{MHz}$ in 512 channels) centered at frequencies of 1366 and 1697\,MHz. A band-switching technique was employed which provided an effective integration time of 6 hours in each band, while retaining the full $uv$ coverage in both. Following calibration, each channel was imaged individually. The Rotation Measure Synthesis (RM-Synthesis) technique \cite{brentjens_debruyn_2005,heald_2009} was used to coherently detect the polarized emission, and its associated RM, across the two observing bands. The key strengths of this method are that it avoids the $n\pi$ ambiguity problem that plagues traditional RM determinations; and that it allows the detection of polarized flux, and its associated RM, even at very low signal-to-noise levels. The typical rms noise level achieved in the WSRT-SINGS polarization maps is $\approx10\,\mu\mathrm{Jy\,beam}^{-1}$. Of the 28 galaxies studied in the polarization survey, 21 were detected in polarized emission. All 21 detections were of spiral galaxies; the few Magellanic and elliptical galaxies in the sample (three and one of these, respectively) were all undetected in polarized emission. The output of the RM-Synthesis step was used to generate maps of polarized intensity, polarization angle, and Faraday rotation measure. Image galleries are presented by \cite{heald_etal_2009}.

\section{Observational Trends and Interpretation}

\begin{figure}[ht!]
\centering
\includegraphics[width=\textwidth]{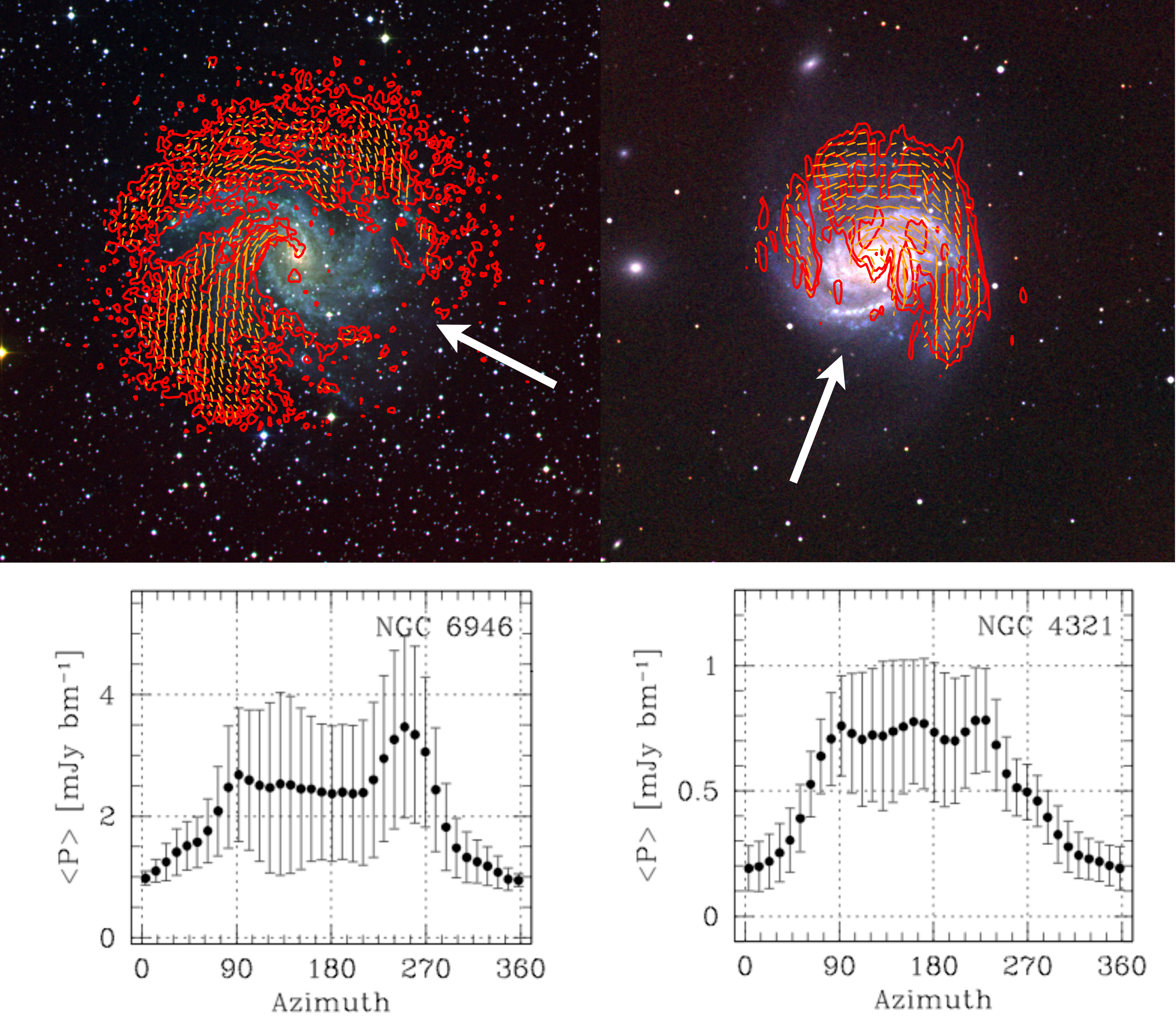}
\caption{Two galaxies observed in the WSRT-SINGS survey: NGC 6946 (left panels) and NGC 4321 (right panels). The optical pictures (top panels) were composed using images from the Digitized Sky Survey. The location of the kinematically receding major axis is indicated with a white arrow. Polarized flux is shown with red contours, starting at $50\mu\mathrm{Jy\,beam}^{-1}$ and increasing by powers of two. Magnetic field orientations, derived from the Faraday rotation corrected polarization angles, are shown with orange lines. The bottom panels show the mean azimuthal variation in polarized flux. The receding major axis corresponds to azimuth $0^{\circ}$.}
\label{fig:n6946n4321}
\end{figure}

\begin{figure}[ht!]
\centering
\includegraphics[width=0.75\textwidth]{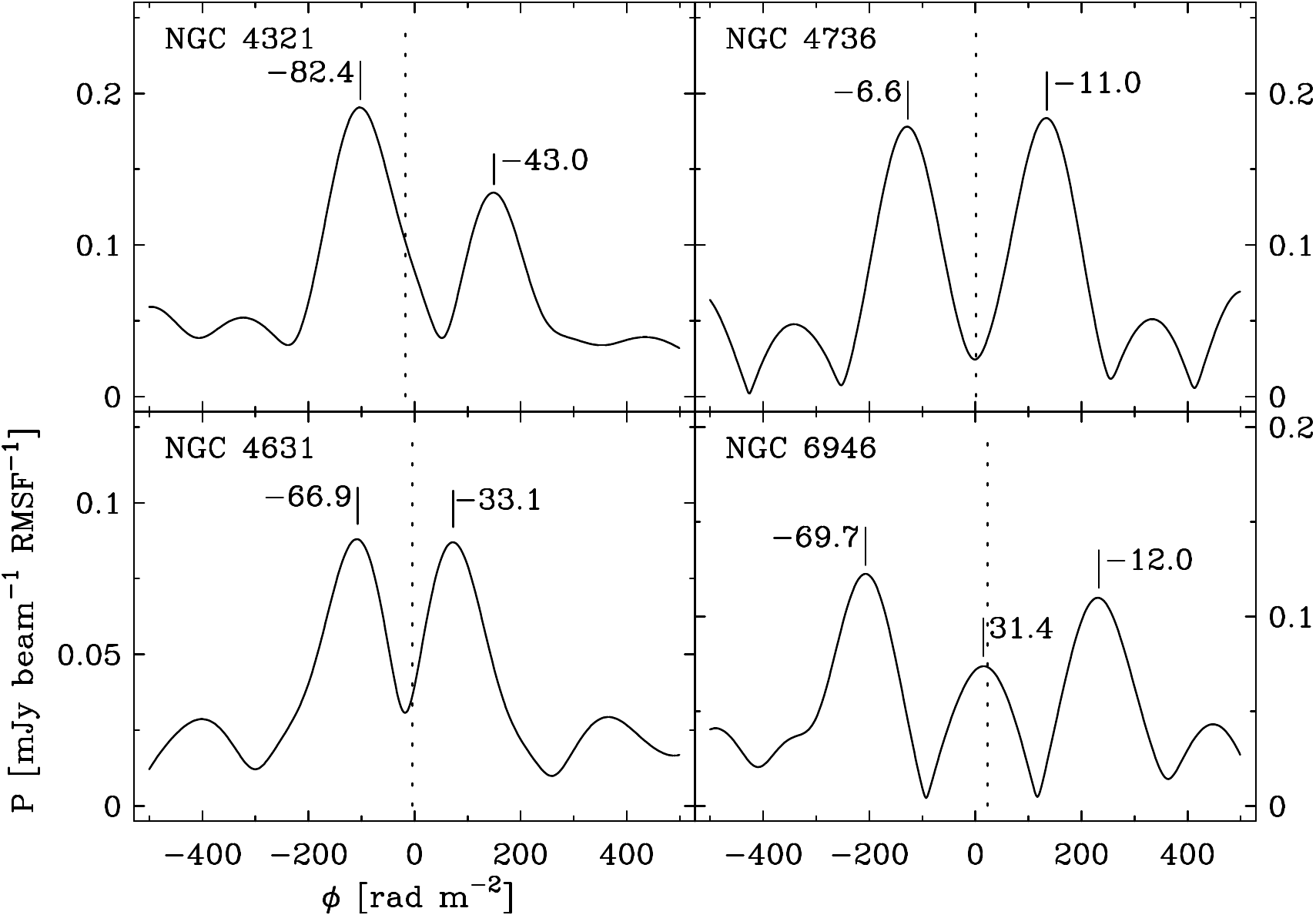}
\caption{Faraday dispersion functions for four of the WSRT-SINGS galaxies, showing the polarized emission as a function of RM in the nucleus of each target. The dotted lines show the estimated local foreground RM value. Polarized emission is clearly detected at two different RM values (perhaps three in the case of NGC 6946), which appear symmetrically distributed about the zero-level. The numbers indicate the polarization angle (in degrees) at the local peak of the Faraday dispersion function.}
\label{fig:nuclei}
\end{figure}

Two major observational trends are observed when considering the resulting data for the entire WSRT-SINGS sample:
\begin{enumerate}
\item All galaxies with extended polarized flux show a clear pattern in its azimuthal distribution. In face-on targets, the minimum in polarized flux always occurs near the kinematically receding major axis. Two of the clearest examples of this behavior, NGC 6946 and NGC 4321, are shown in Figure \ref{fig:n6946n4321}. Highly inclined galaxies have a minimum level of polarized flux along both sides of the major axis, and maxima near the minor axis. The shifting of the peak polarized flux from the approaching major axis to the minor axis appears to occur smoothly as the inclination of the galaxy increases.
\item All galaxies with compact nuclear radio emission have complicated nuclear Faraday spectra. Examples are shown in Figure \ref{fig:nuclei}. The limitations of our observing setup (specifically, the relatively low observing frequencies) make it unclear whether these features are indicative of a detection of multiple Faraday thin media in the galaxy nuclei, or rather a single Faraday thick medium which causes substantial depolarization (at RM values between the main peaks in Figure \ref{fig:nuclei}). Further measurements at higher spatial resolution and higher frequency will be required to clarify the physical situation in these nuclei.
\end{enumerate}

That the polarized flux is always minimized along the {\it receding} major axis in the more face-on targets implies that this feature is a key handle on the fundamental relationship between the structure of the magnetic field, and that of the galaxy itself. In \cite{braun_etal_2009}, we model several three-dimensional forms of magnetic field geometry in an attempt to match the observations of both polarized flux and Faraday rotation measure. A key feature of the models is that midplane depolarization prevents us from observing the backside of galaxy disks at frequencies near 1.4~GHz. In fact, high frequency observations of NGC 6946 (see \cite{beck_2007}) show that the strong azimuthal variation in the polarized intensity vanishes at wavelengths of 3~cm and 6~cm. We interpret this as a reduction in the midplane depolarization at high frequencies. Using this assumption, we are able to constrain the vertical morphology of the large-scale magnetic field, finding that the observations prefer an axisymmetric spiral plus a quadrupolar component. The model is also consistent with the disappearance of the azimuthal dependence of polarized intensity at higher observing frequency. Details of the modeling results are provided by \cite{braun_etal_2009}.

Future observations are needed in order to confirm this trend in a larger sample of galaxies. A larger sample will also provide leverage on any variation in the three-dimensional magnetic field geometry with other galaxy properties such as Hubble type, star formation rate, and rotation speed. Additional high-frequency measurements are also required in order to confirm the prediction that the azimuthal variation in polarized intensity vanishes above $\sim5\,\mathrm{GHz}$.

\section{Future Facilities}

Polarization observations are expected to flourish with the advent of future radio facilities such as the planned WSRT upgrade Aperture Tile in Focus (APERTIF; \cite{oosterloo_tv}) and the Australian Square Kilometre Array Pathfinder (ASKAP; \cite{braun_tv}). These radio telescopes in particular, which are both based on dense Focal Plane Array (FPA) technology, are by their nature expected to provide excellent data sets for making progress in understanding the issues described above. The key desired observational characteristics for future polarization surveys are (1) large field-of-view; (2) excellent control and/or suppression of instrumental polarization; and (3) wide observing bandwidth.

Both APERTIF and ASKAP are expected to perform radio continuum surveys of large areas of the sky; in tandem, it is possible that they will together survey the \emph{entire} sky. The expected noise levels resulting from large-area continuum surveys with both telescopes are comparable to the sensitivity achieved with the WSRT-SINGS survey. With their wide bandwidth, the precision with which RM values can be determined using the RM-Synthesis technique is excellent (for example, using a nominal 1000-1700\,MHz bandwidth, the expected native RM resolution is approximately $\sigma\approx25$\,rad\,m$^{-2}$, and the effective RM resolution improves with increasing signal-to-noise; see e.g. \cite{brentjens_debruyn_2005}). By comparison, the native RM resolution of the WSRT-SINGS observations described above was $\sigma\approx\,61\,\mathrm{rad\,m}^{-2}$.

Polarized radio continuum maps will be one of the data products produced by the polarization surveys to be performed with APERTIF and ASKAP. One of the main science drivers for such a massive polarization survey is the development of an all-sky Rotation Measure grid (RM-grid), which will be a powerful intermediate step as we work towards the incredible RM-grid which will be provided by the Square Kilometre Array (SKA) \cite{gaensler_etal_2004}. During such a survey, a large number of nearby (i.e., spatially resolved) galaxies will also be observed. To roughly estimate the number of resolved galaxies which will be detected in such a survey using APERTIF, the Nearby Galaxy Catalog \cite{tully_1988} was queried for spiral galaxies above a declination of $30^{\circ}$ and with angular size sufficient for $\gtrsim10$ resolution elements at the nominal angular resolution of the WSRT. Such a query results in 226 galaxies. If, as in the WSRT-SINGS sample, $\sim40\%$ have detectable extended polarized emission at 1.4\,GHz, a much larger version of WSRT-SINGS in the northern hemisphere would be able to place about 100 galaxies in a future version of Figure \ref{fig:n6946n4321}. Extending to the entire sky, a huge sample of about 350 galaxies can be built up. Of course kinematic information will also be available for all targets, since H$\,\textsc{i}$ surveys will also be performed by both instruments.

One of the main reasons that FPA designs are expected to be so powerful for large-area polarization surveys is a side-effect of the digital beamforming. To provide such a large field of view, FPAs form multiple, angularly separated but densely packed beams simultaneously on the sky. This is achieved by electronically applying the proper complex weights to the signals coming from the individual FPA elements. It is expected (but remains to be shown with a prototype system) that this same weighting procedure can be used to control the polarization response of the system, with the end result that instrumental polarization can be greatly reduced without a significant impact on the overall sensitivity of the telescope (e.g. \cite{capellen_bakker_2009}).

Of crucial importance in the interpretation of the polarization data obtained for each galaxy is an estimate of the foreground RM contribution. The best way to estimate this quantity is to investigate the RM of polarized background sources surrounding the target galaxy. By utilizing such a ``local RM grid,'' rather than relying on the mean RM value of the target itself, information about a zero-level offset (as might result from a vertical magnetic field in a face-on target) is not lost. The current field of view of the WSRT is sufficient to detect a handful of such surrounding background sources. With future facilities and their large fields of view, the ability to estimate the foreground RM level will be greatly enhanced.

Future radio facilities with smaller instantaneous fields of view, such as the Expanded Very Large Array (EVLA; \cite{rupen_tv}) and the Karoo Array Telescope (MeerKAT; \cite{jonas_tv}) have the benefits of higher sensitivity in a single pointing, and also being able to operate at higher observing frequencies than either APERTIF or ASKAP. They will be able to provide sensitive polarization observations of local galaxies both at $\sim\,1.4\,\mathrm{GHz}$ and at higher frequencies, allowing not only observations of weaker magnetic fields and increased RM resolution due to the increased sensitivity, but also crucial tests of midplane depolarization and the polarization properties of galactic nuclei at the higher frequencies.

\acknowledgments{The Westerbork Synthesis Radio Telescope is operated by ASTRON (Netherlands Institute for Radio Astronomy) with support from the Netherlands Foundation for Scientific Research (NWO).}


\begin{thebibliography}{99}

\bibitem{beck_2007}
R.~Beck, \emph{Magnetism in the spiral galaxy NGC 6946: magnetic arms, depolarization rings, dynamo modes, and helical fields}, \emph{A\&A} {\bf 470} (2007) 539.

\bibitem{beck_2009}
R.~Beck, \emph{Measuring interstellar magnetic fields by radio synchrotron emission}, \emph{IAUS} {\bf 259} (2009) 3.

\bibitem{braun_etal_2007} 
R.~Braun, T.A.~Oosterloo, R.~Morganti, U.~Klein \& R.~Beck, \emph{The Westerbork SINGS survey. I. Overview and image atlas}, {\it A\&A} {\bf 461} (2007) 455.

\bibitem{braun_etal_2009}
R.~Braun, G.~Heald \& R.~Beck, \emph{The Westerbork SINGS Survey. III. Global Magnetic Field Topology}, 2009, submitted

\bibitem{braun_tv}
R.~Braun, \emph{Panoramic Surveys of the Radio Sky with the Australian SKA Pathfinder}, in proceedings of \emph{PRA2009}, \pos{PoS(PRA2009)002}.

\bibitem{brentjens_debruyn_2005}
M.A.~Brentjens \& A.G.~de Bruyn, \emph{Faraday rotation measure synthesis}, {\it A\&A} {\bf 441} (2005) 1217.

\bibitem{capellen_bakker_2009}
W.A.~van Cappellen \& L. Bakker, \emph{Experimental Results of a 112 Element Phased Array Feed for the Westerbork Synthesis Radio Telescope}, \emph{IEEE Symp. on Ant. and Propagat.} 2009.

\bibitem{gaensler_etal_2004}
B.M.~Gaensler, R.~Beck \& L.~Feretti, \emph{The origin and evolution of cosmic magnetism}, \emph{NewAR} {\bf 48} (2004) 1003.

\bibitem{heald_2009}
G.~Heald, \emph{The Faraday rotation measure synthesis technique}, {\it IAUS} {\bf 259} (2009) 591.

\bibitem{heald_etal_2009}
G.~Heald, R.~Braun \& R.~Edmonds, \emph{The Westerbork SINGS Survey. II. Polarization, Faraday rotation, and magnetic fields}, {\it A\&A} {\bf 503} (2009) 409.

\bibitem{jonas_tv}
J.~Jonas, \emph{The MeerKAT SKA precursor telescope}, in proceedings of \emph{PRA2009}, \pos{PoS(PRA2009)004}.

\bibitem{kennicutt_etal_2003}
R.C.~Kennicutt et al., \emph{SINGS: The SIRTF Nearby Galaxies Survey}, {\it PASP} {\bf 115} (2003) 928.

\bibitem{krause_2008}
M.~Krause, \emph{Magnetic Fields and Star Formation in Galaxies of Different Morphological Types}, {\it ASPC} {\bf 396} (2008) 147.

\bibitem{oosterloo_tv}
T.~Oosterloo, \emph{The latest on Apertif}, in proceedings of \emph{PRA2009}, \pos{PoS(PRA2009)006}.

\bibitem{rupen_tv}
M.~Rupen, \emph{The EVLA: Progress and Prospects}, in proceedings of \emph{PRA2009}, \pos{PoS(PRA2009)003}.

\bibitem{tully_1988}
R.B.~Tully, \emph{Nearby galaxies catalog}, Cambridge University Press, Cambridge and New York 1988.

\bibitem{widrow_2002}
L.M.~Widrow, \emph{Origin of galactic and extragalactic magnetic fields}, \emph{RvMP} {\bf 74} (2002) 775.

\end{thebibliography}
\end{document}